\documentclass[twoside,11pt] {article}\textheight 9in\textwidth 6.5in
\usepackage {latexsym,amsmath,amssymb,graphicx,mathptmx,comment}
\begin {document}

\newcommand\K{\mbox{\bf K}}
\newtheorem {theorem} {Theorem}
\newtheorem {lemma} {Lemma}

\newtheorem {proposition} {Proposition}
\newtheorem {corollary} {Corollary}
\newcommand {\ie} {\emph {i.e.,}\ }
\newcommand {\ooo} {\mathcal O}

\newcommand {\KP} {\ensuremath {\mathit \mbox{KP}}}
\newcommand {\zed} {\mathbb Z}

\makeatletter
\def\@yproof[#1]{\@proof{ #1}}
\def\@proof#1{\begin{trivlist}\item[]{\em Proof#1.}}
\newenvironment {proof} {\@ifnextchar[{\@yproof}{\@proof{}
}}{\hfill $\square$ \end{trivlist}}
\makeatother

\title {Complex Tilings\thanks
{Preliminary version of this article appeared as \cite {dls-stoc}.} }

\author {
Bruno Durand\thanks
 {Laboratoire d'informatique fondamentale de Marseille,
39, rue Joliot-Curie, 13453 Marseille, France,
 \hfil\newline \mbox {\texttt {email: Bruno.Durand at lif.univ-mrs.fr}}.}
 \qquad Leonid A. Levin\thanks
 {Boston University, Computer Science department,
111 Cummington St., Boston, MA 02215, USA.
\hfil\newline\mbox {\texttt {email: Lnd at bu.edu}}.
Supported by NSF grant CCR-0311411.}
 \qquad Alexander Shen\thanks
 {Laboratoire d'Informatique fondamentale de Marseille,
 Institute of Problems of Information Transmission, Moscow,
 \hfil\newline\mbox{\texttt {email: alexander.shen at lif.univ-mrs.fr,
shen at mccme.ru}.}\hfil\newline Supported in part by Universit\'e de
Provence, CNRS, Russian Foundation for Basic Research (grants
03-01-00475, 358.2003.1 and others) and the Swedish Foundation for
International Cooperation in Research and Higher Education, grant
Dnr-99/621.}}

\date {\today} \maketitle \begin {abstract}

We study the minimal complexity of tilings of a plane with a given tile
set. We note that every tile set admits either no tiling or some tiling
with $\ooo(n)$ Kolmogorov complexity of its $(n\times n)$-squares. We
construct tile sets for which this bound is tight: all $(n\times
n)$-squares in all tilings have complexity at least $n$. This adds a
quantitative angle to classical results on non-recursivity of tilings --
that we also develop in terms of Turing degrees of unsolvability.

Keywords: Tilings, Kolmogorov complexity, recursion theory

\end {abstract}

\section {Introduction}

Tilings have been used intensively as powerful tools in various fields
such as mathematical logic (see, e.g., \cite {bgg} and references
within), complexity theory~\cite {gurevich91,levin86}, or in physics for
studying quasicrystals (see for instance the review paper~\cite
{ingersent}). In all of these branches the ability of tile sets to
generate ``complicated'' tilings is essential --- it was already clear
in Wang's original papers~\cite {wang61,wang62}.

A \emph {tile} is an unit square with colored edges (each of the four
sides has some color). Assume that a finite set $\tau$ of tiles is
given. We want to form a $\tau$-tiling, \ie to cover plane with
translated copies of tiles from $\tau$ in such a way that adjacent tiles
have a common edge which has the same color in both tiles.

Tiles placed in the plane can be seen as a dual view of crosses on a
grid. A \emph {cross} in a grid is a combination of four (colored) edges
sharing a corner. Given a set of allowed crosses, one may wish to color
all edges of a grid in such a way that all crosses are allowed. This
question is equivalent to the original one. (Turning each edge
orthogonally around its own center turns the grid of edges into its dual
graph and tiles into crosses and vice versa.) Thus, one can use either
representation for best visual advantages.

We call a \emph {palette} a finite set of tiles that can be used to tile
the plane (\emph {+palette} for crosses). The problem to know whether a
set of tiles is or is not a palette, is the so-called \emph {domino
problem}.

In order to prove its undecidability left open in~\cite {wang61,wang62},
Berger~\cite {berger} constructed an aperiodic set of tiles, \ie a
palette $\tau$ such that all $\tau$-tilings are aperiodic (no
translation keeps them unchanged) (see also~\cite {robinson} and \cite
{ad-bgg}).

Hanf in~\cite {hanf} (for the origin constrained case) and then Myers
in~\cite {myers} (for the general case) have strengthened this result
and constructed a palette that has only non-computable (non-recursive)
tilings.

\medskip The aim of this paper is to understand how
``complexity-demanding'' a palette can be, and we measure the complexity
of a palette by the minimal Kolmogorov complexity of tiling it can form.
More specifically, we measure the complexity of regions in the simplest
tiling that can be formed (a formal definition is given in the next
section).

Some information about the complexity of tiling from the
recursion-theoretic viewpoint is also provided (section~\ref
{sec:recurs}).

What can be said about Kolmogorov complexity of a tiling? Tiling is an
infinite object, so we look at $(n\times n)$-squares and measure their
Kolmogorov complexity.

Item~\ref {bound-a} of Theorem~\ref {bound} below states that for each
palette there exists a tiling such that complexity of its $(n\times
n)$-squares is $\ooo(n)$. This bound is tight: item~\ref {bound-b}, our
main result, constructs a palette $\tau$ that has only complex tilings:
in each tiling, every $(n\times n)$-square has complexity at least $n$.
The construction is rather complicated and is based on Berger's
construction in~\cite {berger} and its further developments.

If $\tau$ is a palette with all tilings of at least linear complexity,
then all $\tau$-tilings are aperiodic and non-compu\-ta\-ble because for
every computable (\emph {a fortiori} periodic) tiling, the complexity of
its centered $(n\times n)$-squares is $\ooo(\log n)$.

Note that the right question is ``what is the minimal complexity of a
$\tau$-tiling'' (for a given palette~$\tau$) but not ``what is the
maximal complexity of a $\tau$-tiling''. Indeed, the maximal complexity
could be large for a trivial $\tau$ such as the set of all tiles having
black and white edges where random tiling has complexity $\Omega(n^2)$.

Theorem~\ref {bound} uses the same idea of embedding computations into
tilings that was used to construct palette that has only non-computable
tilings (\cite {hanf,myers}). However, in our case we need embedding
that is ``dense'' enough, and its construction (in non-constrained case)
requires additional efforts.

It seems likely to one of us that Kurdiumov-Gacs hierarchical cellular
automata (see, e.g., \cite {gacs}) could be used instead of Berger's
square hierarchy. This way a stronger result (allowing a constant
fraction of missing tiles in random places) might possibly be achieved.
We, however, are confident in our inability to use these (enormously
complex) constructions and, in any case, the simplicity of the structure
we propose has independent merits.

A more structural approach can be used to handle the complexity of
tilings: a palette can always form ``quasiperiodic'' tilings (see~\cite
{durtcs98}). We can measure the regularity of quasiperiodic tiling by
the growth of the function assigning to an integer $n$ the minimal
``window'' size in which all patterns smaller than $n\times n$ must
appear (for every position of the window and for all patterns that
appear somewhere in the tiling). This approach was further developed
in~\cite {cervdur}.

Theorem~\ref {bound} also implies lower and upper bounds for the number
of different $(n\times n)$-squares in a tiling (Corollary~\ref {count}).

Let us make some remarks about the complexity of tilings from the
viewpoint of hierarchy of Turing-degrees of unsolvability. Hanf and
Myers result in~\cite {hanf,myers} says that some palettes can generate
only non-recursive tilings (\ie of degree $>0$). This cannot be improved
significantly: we cannot find a palette for which all tilings are, say,
$0'$-hard, since for every palette $\tau$, and undecidable set $A$,
there exists a $\tau$-tiling $T$ such that $A$ is not Turing-reducible
to $T$. This result is a corollary of classical results in recursion
theory because the set of all tilings with a given palette is a
$\Pi^0_1$-set (see Odifreddi book p.508~\cite {odifreddi} and references
within\footnote {Thanks to Frank Stephan for references.}).
 However, we present (see Proposition~\ref {recurs}) a short direct
proof of this fact (that goes back to Albert Muchnik and Elena Dyment
and was communicated to us by Andrei Muchnik~\cite {muchnik}).

 \vfil\section {The Main Results}

There are several versions of Kolmogorov complexity $\K(x)$ expressing
``the minimal description length'' (see~\cite {livitanyi}). For our
purposes it does not matter which of them we use.

\begin{theorem}\label {bound}\mbox{}\begin{enumerate}\item\label{bound-a}
 For each palette $\tau$, there exists a $\tau$-tiling in which
 all $n\times n$ squares $n$ have complexity $\ooo(n)$.
 \item\label {bound-b} There exists a palette $\tau$ such that in every
 $\tau$-tiling all $n\times n$ squares have complexity $\ge n$.
 \end {enumerate} \end {theorem}

(The second part can be slightly generalized. A pattern on a connected
subset $P$ of a tiled planar grid is the list of tiles in $P$ and their
coordinates relative to the center of $P$. We encode all objects $x$ in
binary denoting their length as $\|x\|$. In
Theorem~\ref{bound}.\ref{bound-b} we can replace $n\times n$ squares by
patterns of diameter $n$.)

Item~\ref {bound-a} of this Theorem and the following Corollary are
proven in section~\ref {up-b}. The proof of item~\ref {bound-b} involves
several constructions and will be split into several parts. Section~\ref
{fixed-or} considers the easier origin-constrained case. Section~\ref
{enforcement} describes aperiodic tilings used as a background in the
rest of the proof. Section~\ref {organizing} explains the general
structure of the computation embedded in a tiling; Section~\ref
{stripepower} considers computational power of \emph {stripes} (modules
of different ranks that interact during the computation). Finally,
section~\ref {hierarchy} considers the communication between stripes and
explains how all the stripes working in parallel achieve the declared
goal (preventing low-complexity fragments from appearing in the tiling).

 \begin {corollary}\label {count} Let $D_n(\alpha)$ be the number
of different $n\times n$ squares that appear in tiling~$\alpha$. \begin
{enumerate}\item\label {count-a} Each palette $\tau$ has a $\tau$-tiling
$\alpha$ such that $D_n(\alpha)=2^{\ooo(n)}$.
 \item\label {count-b} There exists a palette $\tau$ such that
$D_n(\alpha)\ge 2^n$ for every $\tau$-tiling $\alpha$ and every $n$.
 \end {enumerate} \end {corollary}

(See section~\ref {up-b} for the proof.)

Theorem~\ref{bound} has a finitary version that says, informally
speaking, that (for some palette) a pattern $x$ whose complexity is less
than its diameter $d$ cannot appear in a $N\times N$-square where $N$ is
the polynomial of the time needed to establish that $\K(x)<d$.

Let $U:p\to x$ be the universal algorithm defining Kolmogorov Complexity
$\K(x)$ of $x$ as the minimal length of its co-image $p$. Let $U'(p)=(U(p),
\|p\|)$ enumerate the super-graph of $\K$. Let $T_k(x)$ be the optimal
inversion time of $U'$ on $(x,k)$ and $S_k(x)$ be the minimal space
$U(p)$ needs to compute each digit of $x$ from some $p\in\{0,1\}^k$.

\begin{theorem}\label{finitary}\begin{enumerate}
\item\label{fn-a}
 Every palette $\tau$, for each $N, k\ge \K(N)+O(1)$, has $\tau$-tiled
 $N\times N$ squares $x$\\ with $\K(x)\le k$ and $S_k(x)=O(N)+S_k(N)$.
\item\label {fn-b}
 There exists a palette $\tau$ such that no pattern~$x$ of
 diameter~$d>\K(x)$
 can be extended to a $\tau$-tiled square of diameter $T_d(x)^5$.
 \end {enumerate} \end {theorem}

See Section~\ref{count-fin} for the proof.

\newpage\section {Proof: the Upper Bound (Theorem
                 \ref{bound-a}.\ref{bound-a})} \label {up-b}

\begin {proof}[of the upper bound]

Fix a palette $\tau$. A \emph {border coloring} $b$ of a $(n\times
n)$-square assigns colors to $4n$ tile sides on the square border. A
border coloring $b$ is called \emph {consistent} if it can be extended
to a tiling of the entire $(n\times n)$-square, i.e., there exists a
tiling of $(n\times n)$-square that matches~$b$. Consider the following
algorithm that, applied to a consistent border coloring of a square of
size $n\times n$ where $n=2^k$, extends it to the tiling of the entire
square:

A. Find the alphabetically first coloring of the central lines dividing
the square into four equal squares such that all four squares get a
consistent border coloring.

B. Apply the algorithm recursively to four border colorings of smaller
squares. (For $(1\times 1)$-square consistent border coloring is just a
tile coloring.)

When this algorithm is applied to a square with side $2^k$, it generates
a tree of recursive calls for sub-squares with sides $2^l$ for all $l$
(``standard'' sub-squares). On each standard sub-square the complexity
of tiling is proportional to square side (since tiling is computed by
our algorithm starting from border coloring).

Each non-standard sub-square with side $m$ is contained in $4$ standard
sub-squares with sides smaller than $2m$ and therefore has complexity
$\ooo(m)$. This argument shows that for some constant $c$ and for all
$k$ there exists a tiling of size $2^k\times 2^k$ such that all
$(m\times m)$-sub-squares in this tiling have complexity at most $cm$.

Using compactness argument, we conclude that there exists an infinite
tiling with the same property.

(Here are the details. For a fixed $c$ let us call a tiling of $(n\times
n)$-square \emph {good} if all $(m\times m)$-subsquares of it have
complexity at most $cm$. As we have seen, for some~$c$ there exist good
$(n\times n)$-tilings for arbitrarily large $n$. Call a tiling of
$(n\times n)$-square \emph {extendible} if it appears as a central part
of some good tiling of arbitrary large size. Note that each extendible
tiling can be extended to some extendible tiling of
$(n+2)\times(n+2)$-square by adding one more layer. Continuing this
process, we get a tiling of the entire plane.) \end{proof}

\begin {proof}[of the Corollary]
The first statement of Corollary~\ref {count} is a direct consequence of
upper bound in Theorem~\ref {bound}, since the number of different
objects with complexity $\ooo(n)$ is $2^{\ooo(n)}$.

To prove the second statement we use the lower bound from Theorem~\ref
{bound}. Let $\tau$ be the palette such that all $\tau$-tilings have
complexity at least $5k$ for $(k\times k)$-squares. (We have weaker
bound $k$ in Theorem~\ref {bound}, but it does not matter since we can
combine several tiles into one larger tile.)

Let $\alpha$ be a $\tau$-tiling. Assume that for some $n$ the number of
$(n\times n)$-squares in $\alpha$ is less than $2^n$. Consider a square
of size $k\times k$ where $k$ is a large multiple of $n$, and its
``border'' formed by $(n\times n)$-squares. Since each border square can
be described by $n$ bits (there are less than $2^n$ of them), the whole
border has complexity approximately $4k$ (for large $k$). Then we can
change the tiling, replacing the interior of $(k\times k)$-square by the
alphabetically first tiling that is compatible with the border. Then new
interior is determined by the border, therefore the complexity of new
$(k\times k)$-square is less than $5k$ and this contradicts to our lower
bound (that is valid for all tilings). \end {proof}

\newpage\section {Proof: Origin Constrained Case}\label {fixed-or}

To prove the lower bound, we start with the much simpler ``origin
constrained case''. It means that we consider only those tilings of the
plane that have a fixed tile at the origin. This allows us to enforce
the tiling to be a time-space diagram of a Turing machine.

\subsection {Computation Performed}

Let us agree that each horizontal line in a tiling represents a Turing
machine's tape at time $t$ where $t$ is the vertical position of that
line. The tile used at position $(x,t)$ encodes the contents of cell
number $x$ at time $t$ (including the head state if the head is inside
cell $x$ at time $t$).

The rules of Turing machine are local and therefore can be encoded in
terms of tilings (one may represent overlapping groups of cells on
time-space diagram by a tile). This technique is well known (see, e.g.,
\cite {wang61}) and we won't go into details here. It works only for
constrained case; we require the origin to be the tile that contains
head of TM, otherwise a tiling may contain no computation.

Since tilings can simulate the behavior of a TM with an infinite tape,
it remains to construct a TM that will ensure high complexity of its
time-space diagram.

Imagine that we have a TM with a double tape: each cell is the Cartesian
product of a workspace and an ``input bit''. The TM may change only the
workspace of each cell, the input bit is read-only. This TM can checks
that input sequence is ``complex enough'', that is, its input string has
Kolmogorov complexity at least $ck$ for a constant $c<1$ in each $k$
consecutive columns. More precisely, the set of all finite strings with
low Kolmogorov complexity is enumerable (we can try all the programs in
parallel and look for the cases when the output of a program is
significantly longer than the program itself). Our TM can enumerate such
simple finite strings, compare them with segments of the input tape,
rejecting the tiling if any match is found.

This addresses Theorem~\ref{bound}.\ref{bound-b} for patterns of width
(equal to the number of input bits) close to their diameter.
 For narrow patterns we must superimpose two orthogonal copies of this
construction. This suffices, since diameter of any pattern equals,
within a factor of 2, to either its width or height.

We need now to prove existence of sequences our machine does not reject.

\subsection {Complexity Lemma}

\begin {lemma}\label {complexity}
For each $c\!<1$ there exists a binary sequence $\omega$ with $\K(x)\ge
c\|x\|-\ooo(1)$ for all its substrings $x$. \end {lemma}

\textbf{Remark 1}. For our purposes it is enough to prove this lemma for
some positive $c$ (however small). Still, the more general statement
(for all $c<1$) is of some independent interest, so we prove it in this
stronger form. We cannot strengthen this Lemma further since for each
sequence $\omega$ the above inequality fails for some $c<1$ and an
infinite set of substrings of unbounded lengths. (If every binary string
is a substring of $\omega$, then this is evident. If some string $X$
does not appear in~$\omega$, then the bound $\K(x)\le c\|x\|+\ooo(1)$ is
true for some $c$ and for each substring $x$ of $\omega$.)

\textbf{Remark 2}. It is not important which version of complexity to
use in the lemma since $c$ is not fixed and all versions differ by a
logarithmic term. However, in the proof it is convenient to use prefix
complexity.

\textbf{Remark 3}. In this lemma we speak about sequences that are
infinite in one direction (though the sequence of indices on the tape is
bi-infinite). However, this is not important: if there exists an
infinite in one direction sequence with this property, there are
arbitrary long finite sequences with this property, and the standard
compactness argument shows that there are bi-infinite sequences with
this property.

[In fact, compactness is not even needed here. We can construct a
bi-infinite sequence with this property from a one-way infinite one by
putting bits alternatively left and right at the small cost of a
multiplicative factor 2 on $(1-c)$.]

Thus our lemma proves that the above constructed TM does not halt for
some input sequences (having this property).\footnote {Note that
time-space diagram occupies only the upper half-plane where time is
positive, but this does not matter since high complexity of squares is
guaranteed by input bits which propagate vertically in both directions.
Similarly in extended abstract of this article (STOC~2001~\cite
{dls-stoc}) the complexity of all squares was assured by each square
propagating its input bits vertically, horizontally, and diagonally.
(The last direction of propagation was among many details missing in
that abstract. It referred to the version posted at arXiv.org
simultaneously with STOC~2001 for more details; now, we give an entirely
different, and simpler, proof of a stronger result.)}

\begin {proof}[of Lemma~\ref{complexity}]

Let us prove first for some constants $c',c''$ that for every string $x$
and natural number $n$ there exists an $n$-bit string $y$ such that
   $$ \KP(xy)+\KP(n)+c'\ge \KP(\langle x,y\rangle)+c''\ge \KP(x)+n
   $$ Here $\langle x,y\rangle$ stands for the encoding of the ordered
pair formed by $x$ and $y$. The second inequality is true since of $2^n$
such pairs with a given $x$ some must have a universal semimeasure
smaller than $x$ at least by a factor of $2^n$ and thus an $n$ bits
higher complexity.

And the first inequality is true since $\langle x,y\rangle$ can be
reconstructed from $xy$ and $\|y\|=n$.

Now we can prove the lemma as follows. For a given $c<1$ we choose $m$
such that
   $$ m-\KP(m)-c'\ge cm.
   $$ Then, starting with an empty sequence, we add blocks of length $m$
to it in such a way that each block increases the complexity at least by
$cm$. Adding several blocks, we increase the length by some $M$ (which
is a multiple of $m$) and the complexity at least by $cM$. Since
   $$ \KP(uv)\le \KP(\langle u,v\rangle)+O(1)\le \KP(u)+\KP(v)+c''',
   $$ the group of added blocks has complexity at least $cM-c'''$. Thus
we have proved our Lemma for segments that start and end at coordinates
that are multiples of $m$. The boundary effects can be compensated by a
small change in $c$. \end {proof}

\section {Proof: Self-similar Pattern}\label {enforcement}

Now we have to consider the general case, no more requiring a fixed tile
at the origin. Let us start with some informal remarks. The palette must
prevent individual tilings from being periodic. This can be provided by
a ``self-similar'' structure: tiling is divided into ``mega-tiles''~---
blocks of large sizes (squares of size $2^n\times 2^n$ for all $n$) that
behave like individual tiles.

A self-similarity of this type was used in Robinson's construction of an
aperiodic tiling (see Robinson's original paper~\cite {robinson} and an
exposition given in~\cite {ad-bgg}). A slightly more rigid construction
(where all ``mega-tiles'' are aligned, which was not the case for
Robinson's palette), is explained in~\cite {levin04} and~\cite
{dls-mathint}. We do not repeat the latter construction here but just
describe the self-similar pattern that can be enforced by it.

\begin {figure}[ht]
\hfill\parbox[b]{3.2in}{
 $$\includegraphics[scale=0.5]{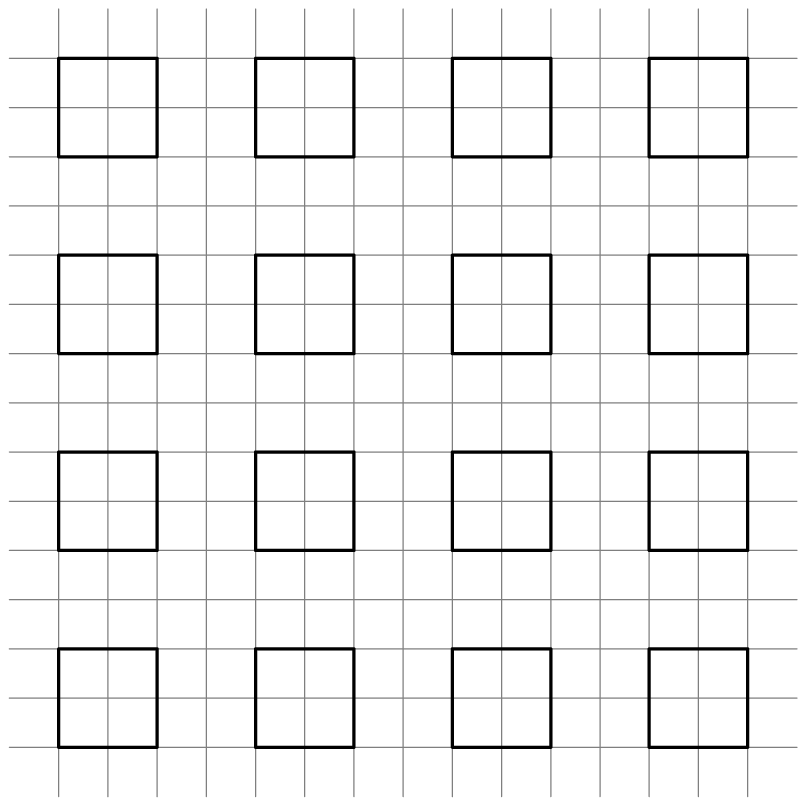}$$ \caption {Rank 1 squares
\label {self.1.eps}}
 }
 \hfill\parbox[b]{3.2in}{
 $$\includegraphics[scale=0.5] {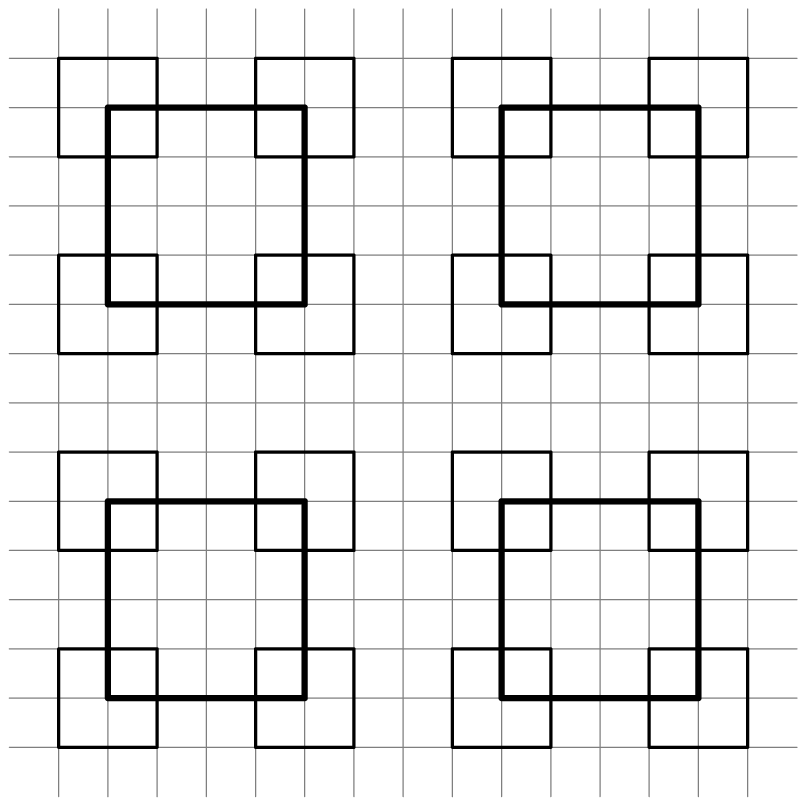}$$ \caption {Rank 2 squares
\label {self.2.eps}}
 }\hfill
 \end {figure}

Consider a grid of $2\times 2$ squares separated by two cells (Fig.~\ref
{self.1.eps}).

Then group these squares into groups of four squares whose centers form
a twice larger square (Fig.~\ref {self.2.eps}). We get a rank 2 grid
formed by $(4\times 4)$-squares separated by four cells; this grid is
twice larger than the rank 1 grid.

\begin {figure}[ht] $$\includegraphics[scale=0.5]{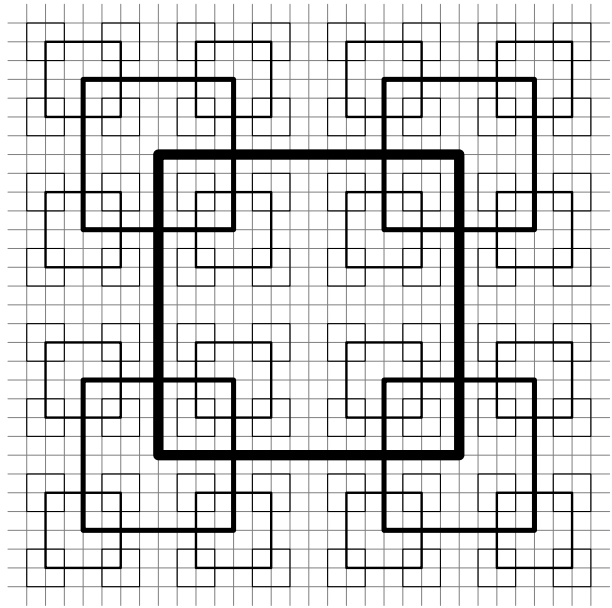}$$
\caption {Hierarchy of squares} \label {self.3.eps} \end {figure}

Then we group rank 2 squares into groups of four squares whose centers
are corners of rank 3 square etc. We will refer to the edges on the
borders of squares of all ranks as \emph {dark}. (Fig.~\ref {self.3.eps}
shows one rank 4 square and underlying hierarchy of smaller squares.)

Note that we have described not one specific pattern but an uncountable
family of patterns: at each rank we have a two-bit choice while grouping
squares into 4-groups. Therefore, a pattern (together with a specified
cell in it) is determined by an infinite sequence of bits.

Note also that the dark pattern may be either connected or not. The
latter happens if there exists a ``separating line'' (a vertical or
horizontal line that does not intersect any dark square). The dark
pattern in these cases has $2$ or $4$ connected components separated by
either a horizontal or vertical line, or by an infinite cross. For the
case of one separating line we require it to be of uniform color (either
light or dark). For the case of two separating lines they must form a
dark corner (four possible orientations).

Looking at some non-separating line, we see that it consists of
alternating dark and light segments of length $2^{k+1}$ where $k$ is the
rank of dark squares adjacent to it. We say that this line has rank $k$
(is a $k$-line).

We assign infinite rank to separating lines. The non-connected case
(when separating lines exist) is called the \emph {degenerated case} in
the sequel and requires special treatment, see Subsection~\ref
{complexity-check}.

\begin {proposition}\label {self-similarity-proposition}
There exists a palette and a projection of its colors into
$\{\text{dark},\text {light}\}$ such that every tiling is projected onto
some pattern of the described type. \end {proposition}

See~\cite {levin04} and~\cite {dls-mathint} for the proof. This palette
$P$ (+palette, actually) provides two \emph {orientation bits} which
will help us below. These bits on each edge $x$ show vertical and
horizontal direction to the nearest center of the dark square with
border co-linear to $x$. These bits form distinct crosses at the
intersections of lines of the same rank, of adjacent ranks, and of more
distant ranks. In a more informal language (that will be used in the
sequel) one can say that each edge on a dark square ``knows'' (\ie this
information is encoded in its color) whether it belongs to the left or
to the right half of the square. The same is true for the lower and
upper half of the square (and vertical edge). The corner node ``knows''
(\ie this information is encoded in the neighbor colors) that it is a
corner node, etc.

\begin {figure}[ht]
 \hfill\parbox[b]{3.2in}{ $$\includegraphics[scale=1]{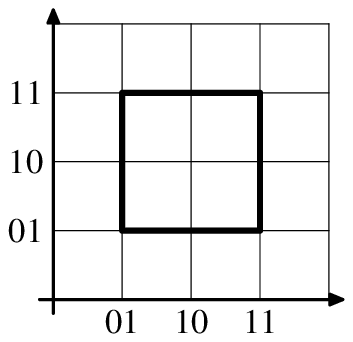}$$
\caption {$4$-coordinates and rank~$1$ squares \label {coord.1.eps}}
 }
 \hfill\parbox[b]{3.2in}{ $$\includegraphics[scale=1]{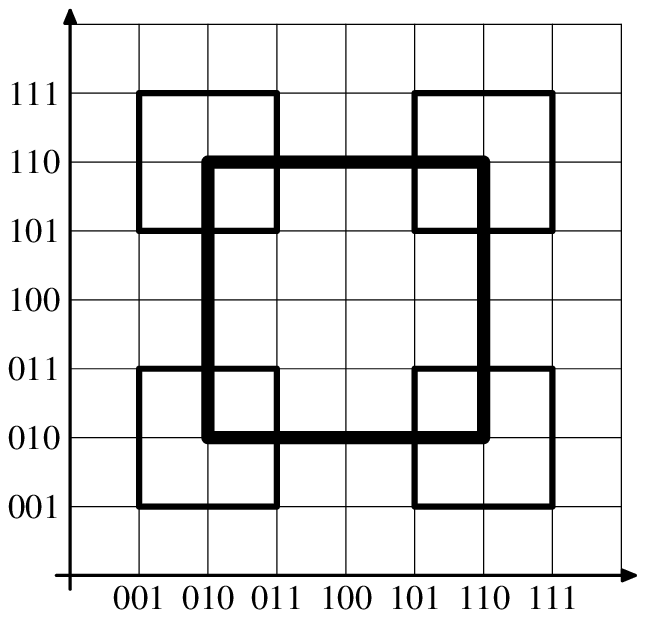}$$
\caption {$8$-coordinates and rank~$2$ squares \label {coord.2.eps}}
 }\hfill
 \end {figure}

To provide more formal description of the pattern, it is convenient to
use a kind of $2$-adic coordinates. Consider lines that go in-between
$2\times 2$ squares (one line per four cells). Taking them as reference
lines, we provide ``modulo $4$'' coordinates, or just $4$-coordinates,
as shown in Figure~\ref{coord.1.eps}.

The same can be done modulo~$8$ for rank~$2$ squares, as shown in
Figure~\ref{coord.2.eps}.

Note that $8$-coordinates are consistent with $4$-coordinates (two last
bits of the $8$-coordinate form the $4$-coordinate). We can then
consider $16$-coordinates, $32$-coordinates etc. They extend each other,
and for each point of the grid we get a $2$-adic coordinate that is an
infinite (to the left) sequence of bits.

Vertical sides of rank $k$ squares have $x$-coordinates
$\ldots10^{k-1}$; the same is true for $y$-coordinates of horizontal
sides. So we can assign rank to vertical and horizontal lines ($1$ plus
the number of zeros at the end of their coordinates). Lines of rank $k$
contain sides of rank $k$ squares. Each line has some uniquely defined
rank, except for the line with zero coordinates. This line can exist or
not depending on the pattern (this is the separating line we mentioned
above).

The centers of rank~$k$ squares have coordinates that end with $k$
zeros, i.e., lie on the rank $(k+1)$ lines. Other features of the
pattern can be also easily expressed in terms of coordinates. For
example, a vertical grid line with coordinate~$x$ intersects (the
interior of) rank $k$ squares if and only if last $k+1$ digits of~$x$
belong to the open interval $(010^{k-1}, 110^{k-1})$.

\newpage\section {Proof: Stripes and Grids}\label {organizing}

Let us start with some informal remarks. The high complexity of tilings
comes from an \emph {input sequence} $I$, horizontal and infinite in
both directions. Each bit occupies a vertical line. A Turing Machine
(TM) verifies that $I$ has no low complexity segments. This computation
represented by tilings as space-time diagrams.

As it was done for the origin-constrained case, the configuration of a
TM (the contents of its tape including the head position) is represented
by colors of horizontal edges. Their shifts in the vertical direction
represent time evolution. The consistency of states in subsequent
moments of time is achieved via vertical edges that carry the state
information from one horizontal line to another. The correctness of
state transitions is assured by a palette that restricts the coloring of
crosses of these vertical and horizontal lines.

In the origin-constrained case the whole tiling represented one
computation. Now, instead, we arrange infinitely many coexisting and
interacting computations. It is not a problem to combine two
computations at the same location: the Cartesian product of two finite
alphabets is still finite. But this cannot be done for infinitely many
computations. Instead, they are separated in space and time so that each
edge is used only by a limited number of them.
 All the computations should then communicate with each other to check
that every substring of the input sequence has high complexity.

\begin {figure}[ht] $$\includegraphics[scale=1]{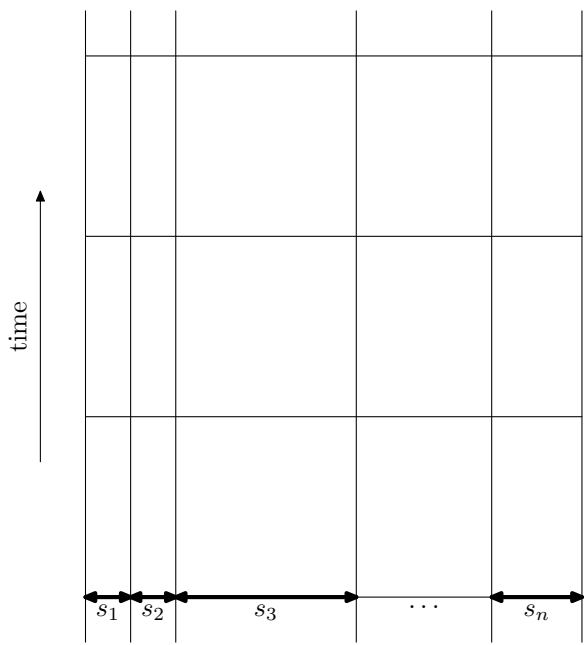}$$
\caption {Subgrid for one computation} \label {self.4.eps} \end {figure}

The organization of these processes ``formats'' the plane using the
self-similar Block Pattern (described above). This formatting is used to
arrange space for infinitely many ``computations''. Each computation is
performed by a \emph{subgrid} that consist of finite number of
(infinite) vertical lines and infinite number of (finite) horizontal
lines arranged as in Fig.~\ref{self.4.eps}. Their intersection points
are called \emph{nodes} of the subgrid.

 Each horizontal line is divided by nodes into segments
($s_1,\ldots,s_n$). Each segment carries one symbol of TM tape or the
state of one cell in the cellular automaton. The changes happens in
nodes only (both for horizontal and vertical lines). All this can be:

\textbullet\ organized locally if each edge of the subgrid knows its
place in the subgrid (whether it is at the node, lies on the left/right
boundary, or between nodes, etc.).

\textbullet\ used in a usual way to simulate computations of TM (or
cellular automata) with the tape of fixed size.

Note that the ``physical'' distances between the grid lines can be
arbitrary, they do not affect at all the computation performed on the
grid. (In fact all the vertical distances will be the same, but not the
horizontal distances. This is somehow shown in Figure~\ref{self.4.eps}.)

Note also that the edges included in the grid do not know how far they
are from the nodes they connect (it is not needed and also there is not
enough colors to encode this).

\medskip

\begin{figure}[ht]
 \hfill\parbox[b]{3.2in}{
 $$\includegraphics[scale=1]{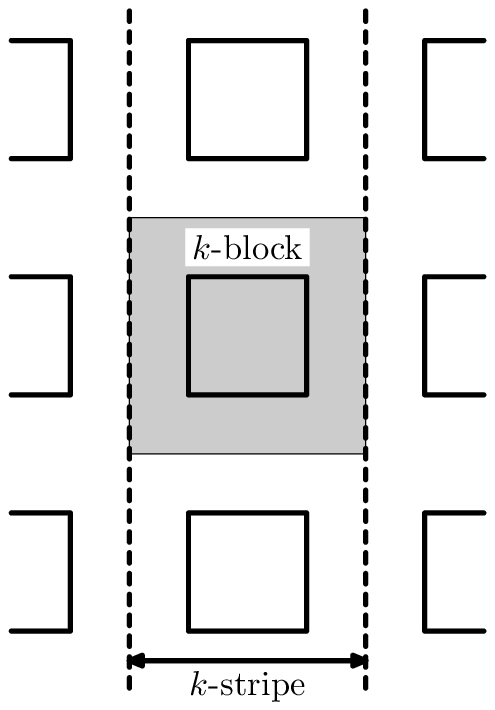}$$ \caption {Grid of $k$-blocks
and $k$-stripe \label {self.5.eps}}
 }
 \hfill\parbox[b]{3.2in}
 {
 $$\includegraphics[scale=1]{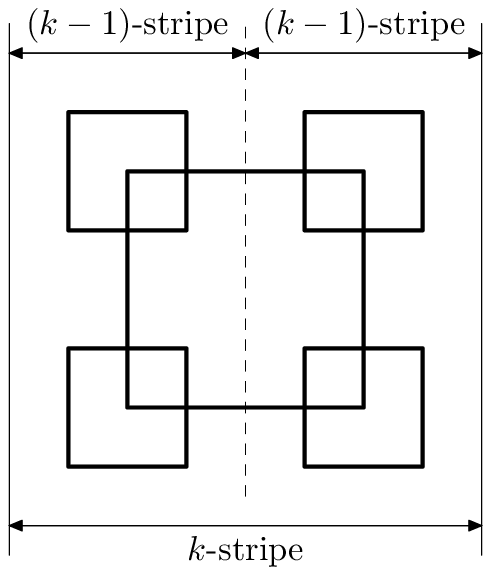}$$ \caption {Stripe and its two
children \label {self.6.eps}}
 }\hfill
 \end{figure}

Now we describe how the subgrids are localized. Let us introduce some
terminology. Each dark square of rank $k$ is included in a twice larger
\emph {$k$-block}. For each $k$, the plane is split into $k$-blocks. A
bi-infinite column of vertically aligned $k$-blocks forms a \emph
{$k$-stripe} (see Fig.~\ref{self.5.eps}).

The borderlines between $k$-blocks have $2$-adic coordinates that end
with $k+1$ zeros.

Each $k$-stripe is a union of two $(k-1)$-stripes, called its \emph
{children} (see Fig.~\ref{self.6.eps}). (For example, $3$-stripe
$[\ldots00000,\ldots10000]$ that lies between vertical lines with
coordinates $\ldots 00000$ and $\ldots 10000$ is the union of two
$2$-stripes: the left is $[\ldots 00000,\ldots 01000]$, the right is
$[\ldots 01000,\ldots 10000]$.

 \vfil

We provide for each stripe a subgrid that is located inside the stripe.
(This subgrid hosts a computation that is finite in space but infinite
in time.) These computations are not completely independent: each
computation communicates with its parent computation (located in the
parent stripe) and its two children computations (located in children
stripes). The communication is implemented by sharing some vertical
lines, \emph{communication} lines. There are three communication lines
in each subgrid: the center line and the vertical lines that contain the
vertical side of $k$-square; the latter two lines are the leftmost and
rightmost lines in the subgrid and are called ``border lines''.

Note that the center line of $k$-square is at the same time the border
line for the $k+1$-level subgrid. Using the center line, $k$-computation
can communicate with its parent $(k+1)$-computation, and using the
border lines, $k$-computation can communicate with its two children
$(k-1)$-computations.

\begin {figure}[ht] $$\includegraphics[scale=1]{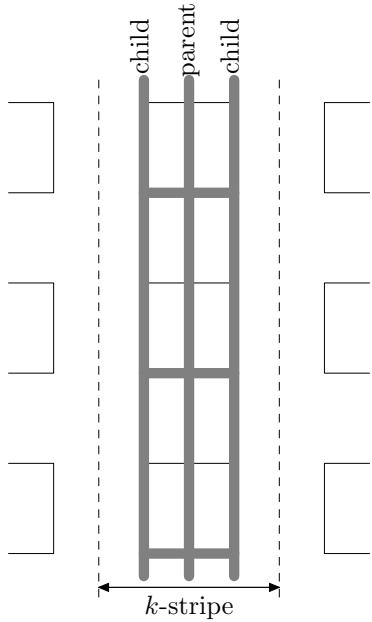}$$
\caption {Horizontal and some vertical subgrid lines for a $k$-stripe}
\label {self.7.eps} \end {figure}

We need also to specify the other vertical lines that are included in
the $k$-level subgrid. They are called \emph{$k$-channels} and lie
between the border lines (see below about their exact location). The
horizontal lines of the subgrid, called \emph{$k$-tapes} in the sequel,
are just bottom lines of $k$-squares (see Fig.~\ref{self.7.eps}).

\section{Designation of $k$-channels}

Designating $k$-channels, we should have in mind that:

(1)~Each vertical line should be shared by a limited number of
computation subgrids (in fact, three in the construction explained
below: it can be a communication line in two subgrids and a channel in
the third one).

(2)~There should be sufficiently many $k$-channels to provide enough
space for the computation performed by a subgrid. (In our construction
the $k$-level subgrid has $\Theta(2^{k/2})$ vertical lines, which is
about the square root of its geometric size.)

(3)~Each node should know its role in every subgrid it belongs to.
(Since the number of subgrids is limited, this is a finite amount of
information that can be encoded in the finite number of colors. What is
important, we need the correct information be enforced by local rules.)

Let us explain how all three goals can be achieved. First, let us agree
that we have two types of dark squares, say, \emph{red} and \emph{blue}
squares, and the colors alternate (so $2k$-squares are red and
$(2k+1)$-squares are blue or vice versa). This is very easy to arrange:
an additional bit distinguishes between red and blue squares, and this
bit should differ for two intersecting dark squares (note that if two
squares intersect, their ranks differ by~$1$).

Then for a vertical line we define its color (red or blue) as the color
of the square whose side it contains. Since the vertical sides of a
$k$-square have coordinates $\ldots010^{k-1}$ and $\ldots110^{k-1}$, the
color of a vertical line depends on whether the number of trailing zeros
is even or odd. (The color is defined uniquely for all lines except for
the separating vertical line if it exists in the pattern; the separating
line can have arbitrary color.) Note that we can easily distribute the
color along the line, so each point on the line knows the line color.

Now the rule:
 \begin{quote}
for each line $L$ we look for the smallest square of the same color that
intersects $L$ (not taking into account the square that has $L$ as its
border); if this square is of rank $k$, the $L$ line is declared to be
$k$-channel and belongs to the computational subgrid for the
corresponding $k$-stripe. \end{quote}

In other terms, consider a line of rank $k$. It is a border line for
$k$-squares. The construction guarantees that it is a center line for
$(k-1)$-squares and does not intersect smaller squares. So we should
look for squares of rank $k+2$, $k+4$ etc. until we find a square that
intersects this line. (Note that rank $k$ line cannot go through the
center or along the sides of those squares.)

\begin {figure}[ht] $$\includegraphics[scale=1]{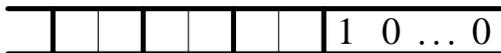}$$
\caption {Finding the square} \label {coord.3.eps} \end {figure}

For $2$-adic coordinates: first we find the rank of the line looking at
trailing zeros. Rank $k$ means that there are $k-1$ trailing zeros,
i.e., the coordinate ends with $10^{k-1}$. Then we split bits on the
left into $2$-bit blocks, as shown in Figure~\ref{coord.3.eps}. Going
from right to left, we find the first block that contains $01$ or $10$
(this means that the line intersects the corresponding square).

So we see that goal~(1) is achieved.

To estimate the number of $k$-channels in a given $k$-subgrid, we can
either use the coordinate description or geometric argument. The
coordinate description shows that we can use blocks $00$ or $11$ after
the $01$/$10$ block and the trailer of the form $1\ldots 0$. This gives
$2^{k/2+O(1)}$ options for $k$-squares. (For each of $k/2+O(1)$ levels
we exclude two of four possible blocks $00$, $01$, $10$ and $11$, i.e.,
half of the lines.)

\begin {figure}[ht] $$\includegraphics[scale=1]{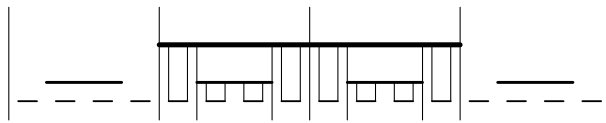}$$
\caption {Grandchildren and their grandchildren} \label {coord.4.eps}
\end {figure}

We can count also the $k$-channels for a given $k$ in a top-down
fashion. Each $k$-square has two $(k-1)$-children of the opposite color
and four $(k-2)$-grandchildren of the same color, see
Figure~\ref{coord.4.eps}. (This figure does not keep the vertical
distances since only horizontal positions matter now.) Two grandchildren
(and all their descendants) lie outside the zone of $k$-square (i.e., on
the left or on the right of $k$-square). Two other provide four
$k$-channels that are lines of rank $k-2$.

Each of these two grandchildren has four grandchildren; two of them are
shadowed by $(k-2)$-squares (and produce $(k-2)$-channels, not
$k$-channels); each of two other produces two $k$-channels, so we get
$8$ $k$-channels that are lines of rank $k-4$.

We continue by induction and conclude that our $k$-square has $4^t$
descendants of rank $k-2t$; some of them are shadowed by squares of
intermediate level (together with the whole stripe), some lie outside
the zone of the initial $k$-square (together with the whole stripe), and
$2^{t}$ (together with the whole stripe) are not shadowed and lie inside
the zone (together with the entire stripe) producing $2^{t+1}$
$k$-channels being lines of rank $k-2t$.

So the total number of $k$-channels in the zone of some $k$-square is
$4+8+16+\ldots$, and this sum has $k/2+O(1)$ terms, so it is equal to
$\Theta(2^{k/2})$. The goal (2) is achieved.

For (3), let us look at some vertical line. It knows its color. Also the
points where it intersects the squares of the same color are locally
known. We can consider them as ``brackets'' (opening bracket means that
line comes in the square, and closing bracket means that is goes out).
So we reduce our task to the following problem: having a correct bracket
structure on a line, find the innermost brackets. It is easy to do by
local rules, if each brackets sends a signal in the outside direction:
the innermost bracket is the bracket that does not receive that signal.

\newpage\section {Proof: Computational Subgrids and their Power}
   \label {stripepower}

Looking at a computational subgrid, we may ignore the rest of the plane
as well as the geometric parameters of the embedding of this subgrid
(distances between the lines etc.). For us it is just a vertical stripe
that obeys some local rules (rules for the left/right boundary could
differ).

Such a tiling represents the time-space diagram of a finite cellular
automaton (finite number of cells that change their state depending on
the previous states of themselves and of their neighbors, according to
some rule which is the same for all automata except for the leftmost and
rightmost one, who have special rules).

This confronts us with the problem we started with: how to initiate a
computation? But now the situation is different: we work in a stripe of
a finite size, so the leftmost and rightmost cells know that they are on
the boundary, and this allows us to restart computation ``once in an
exponential while'' using a counter.

This type of self-stabilization is easy to achieve. We may simulate a
time-space diagram of a Turing machine. To ensure that a head of TM
exists and is unique, we may require that some bit is $0$ at the left
end, is $1$ at the right end, is monotonic (local rule) and the place
where the bit changes behaves like a head of TM. This machine can
perform counting in a positional number system adding $1$ to the counter
all the time. When an overflow happens, we have to restart some other
computation simulated in parallel by the stripe. If the base of our
positional number system is large enough, the counting process takes
more time that the other computation we simulate (if the latter
computation does not repeat itself).

Let us illustrate this technique using the model example of an isolated
stripe.

Consider a vertical stripe (finite in horizontal direction and
bi-infinite in vertical (temporal) direction. Assume that left and right
borders of the stripe have special ``left'' and ``right'' colors. We
want to tile the stripe with tiles from a given tile set (respecting the
border colors).

Let us assume also that all tiles of our palette are divided into two
types ($0$-tiles and $1$-tiles) and tiling rules guarantee that all
tiles on a vertical line have the same type. This guarantees that each
vertical line in a tiling carries one bit (its type) and thus each
tiling of stripe of width $w$ determines a bit string of length $w$.

So for each palette $\tau$ we get a set of strings $L_\tau$ that
corresponds to all $\tau$-tilings of stripes (strings of length $n$ in
$L_\tau$ correspond to the tilings of the stripe of width~$n$).

\begin {proposition}\label {pspace} \begin {enumerate}
 \item\label {pspace-a}
 For every tile set $\tau$ the language $L_\tau$ belongs to PSPACE.
 \item\label {pspace-b} Every language $L$ that is decidable in linear
space can be represented as $L_\tau$ for some tile set~$\tau$.
 \end {enumerate} \end {proposition}

\begin {proof}
Part~\ref{pspace-a} can be proved in the same way as Savitch theorem
($\text{NPSPACE}=\text{PSPACE}$). A $\tau$-tiling of an infinite stripe
exists if and only if there exists a tiling of a
 $w\times h$ rectangle (where $w$ is stripe width and $h\in
2^{\ooo(w)}$) that has the same colors on top and bottom lines. Indeed,
rectangles with this property can be combined into a (periodic) tiling
of the entire stripe. On the other hand, there is only $2^{\ooo(w)}$
possible colorings in a horizontal section, so in every tiling of an
infinite stripe identical sections appear at distance at most
$2^{\ooo(w)}$. Now we can write a recursive procedure that checks
whether there exists a tiling of a rectangle of width $w$ and height $h$
with given top and bottom that runs in space $(w\log h)^{\ooo(1)}$ and
use it to determine whether a given string of length $w$ belongs to
$L_\tau$. (Consider sequentially all possibilities for the colors along
the middle line of the tiling and for each possibility make two
recursive calls for the two parts of the tiling.)

Part~\ref {pspace-b} uses self-synchronization technique explained
above: since every computation with space $\ooo(w)$ terminates in time
$2^{\ooo(w)}$, we can superimpose the computation with a TM computation
that keeps a $\ooo(w)$-bit integer (counter) in each horizontal line and
increases it by $1$ all the time. When an overflow happens, the main
computation is ``rebooted'' in that line. As we have mentioned, this
rebooting procedure gives the main computation enough time to terminate.

This proposition is not directly used in the proof of main theorem. It
is presented here as an illustration of the self-stabilization technique
used in in our main construction together with other key ingredient, the
hierarchical communication between the stripes. \end{proof}

\section {Proof: Hierarchy of Computations}\label {hierarchy}

We have seen that a subgrid can compute every predicate that can be
computed in linear (in its effective width: $\Theta(\sqrt{N})$ for
square with side $N$) space.

However, our construction of tiling with linear complexity of squares
will use subgrids in a more subtle way: different computations
communicate with each other and therefore become parts of some global
computation. Let us explain how it is done.

\subsection {Input Bits}

We assume (as we have done in the constrained case) that each vertical
line carries one bit (that propagates vertically). Therefore, tiling
determines a (horizontal) bi-infinite sequence of bits. Our goal is to
guarantee that this sequence does not contain substrings with low
Kolmogorov complexity (in the same way as for origin-constrained case).

\subsection {Zones of Responsibility}

Recall that computation subgrid based on rank $k$ squares was located in
the middle part of a stripe twice wider than the squares themselves.
This stripe contains $2N$ vertical lines (and input bits) and we say
that these lines and bits are in the ``zone of responsibility'' of this
subgrid. Figure~\ref{self.6.eps} shows $k$-stripe that is the zone of
responsibility for a computational subgrid based on the large square in
the middle; this $k$-stripe is the union of its two children who are
$(k-1)$-stripes. These stripes are zones of responsibility for two
computation subgrids based on the smaller squares.

For each $k$ the entire plane is divided into non-overlapping
$k$-stripes that are zones of responsibility for subgrids based on
$k$-squares. Each zone is divided into two children who are zones of
responsibility for smaller subgrids; these zones are then divided into
smaller zones etc. So we get a tree-like structure of degree $2$ whose
vertices are computation subgrids. Each has two children and one parent,
except for the smallest ones that do not have children.

\subsection {Communication Between Stripes}

Communication between parent and child subgrids is easy to organize
since parent and child share some line that can be used as a meeting
point for the corresponding computations (Turing machines). Of course,
the heads of two machines need not to be at the same time at the meeting
point, and this creates some delay.

Note that visibility of finite number of bits is enough for asynchronous
communication (one of the bits can be used as ``ready'' flag while other
are used as information bits), and the delay ($\ooo(\sqrt{N})$ steps for
each transaction, if TM visits regularly the meeting points) is
acceptable (see below). Therefore we assume that serial asynchronous
communication between parent and its children is possible.

\subsection {Bit Servers}

It remains to explain what each subgrid computes. It runs in parallel
(for example, using time sharing --- or we can simulate two-head TM) two
processes. The first process is called BitServer; the second one is
called ComplexityCheck. Let us explain first what BitServer does.

It serves requests about bits in the zone of responsibility of the
subgrid where it runs. Such a request contains bit address (relative to
the start of the zone) of logarithmic length and should be answered by
providing a value of the input bit with given address. BitServer uses
the most significant bit of the address to determine to which child it
should forward the remaining part of the address, and then waits until
this child provides a reply (which is then sent to BitServer's client).

This recursion stops at the lowest rank where each of $\ooo(1)$ required
input bits is at $\ooo(1)$ distance from the computation, so we may
assume that these bits are directly accessible.

Note that BitServer is able to provide inputs bits from the entire zone
of responsibility (even outside of the physical location of the subgrid)
due to its children.

\subsection {Complexity Check}\label {complexity-check}

BitServers are useless if nobody uses them, so we need to describe the
second process, ComplexityCheck. This process runs in each subgrid and
tries to check whether the sequence of input bits in its zone of
responsibility has no substrings with small Kolmogorov complexity (as it
was done for origin-constrained case). The ComplexityCheck process gets
bits from the BitServer of the same computational stripe. BitServer
interleaves requests from ComplexityCheck with external requests (see
``Time bounds'' below).

The problem is that (compared to the origin-constrained case) each
computation stripe has limited abilities: it can perform only
$\sqrt{N}$-space computations to check $\ooo(N)$ bits. Therefore, each
stripe should rely on higher ranks for complete check (the computation
time needed to find that some string has low Kolmogorov complexity is
not bounded by any computable function of string's length).

This cooperation between ranks is indeed possible if ComplexityCheck is
organized in a proper way. This process generates the list of
``forbidden'' string (strings with low complexity). Thus all ranks
compute the same list in the same order (computation terminates when it
meets time/space constraints). When a forbidden string appears in a
computation it is tested against all substrings of the same length that
are in the responsibility zone of the involved computation stripe. (This
testing is performed by requiring bits from BitServer.) If a forbidden
string is found in the input sequence, then the computation halts
(making the tiling impossible).

If some string has low complexity, then it appears in the list of
forbidden strings and therefore the computational stripes of
sufficiently high rank will have enough time first to generate it and
then to check all substrings against it.

Can we conclude now that all the substrings in the bi-infinite sequence
of input bits have high complexity because they are ultimately checked
in the computational stripes of all ranks? No. The problem is that the
tree-like structure of computational stripes may consist of two disjoint
parts that never meet. (This is a ``degenerate case'' discussed
earlier.) But this does not hurt us because in this case each substring
of the input (in the worst case) consists of two parts that are checked
separately. Taking the longer part, we see that it is not forbidden and
has high complexity. Therefore, every (sufficiently long) substring of
input sequence has complexity at least $cn$ where $n$ is its length and
$c$ is some constant. We can then increase $c$ by combining several
tiles into one bigger tile.

\subsection {Time Bounds}

The only thing that we still have to check is that all this
communication and computation can be performed in $\sqrt{N}$-space (and
$\exp(\sqrt{N})$-time). Indeed, each bit address takes logarithmic space
(and this is much smaller than $\sqrt{N}$-space that is available). The
depth of recursion is logarithmic. So if we assume that BitServer
interleaves internal requests (from ComplexityCheck of the same rank)
with external requests, the time to fulfill them will be still
$\exp(\ooo(\log))$, i.e., polynomial. Note also that the slowdown
induced by testing a forbidden string (when it appears) against all
substrings in the zone of responsibility is polynomial (in the width of
the zone), while the time bound is exponential. So this slowdown does
not prevent the generating process from generating every forbidden
string (at a high enough rank).

This ends the proof of Theorem~\ref {bound}.

\section{Proof: Theorem \ref {finitary}}\label {count-fin}

\begin {proof}
 To prove Theorem~\ref{finitary}.\ref {fn-a}, denote $s_N$ the space
used by the algorithm of Section~\ref {up-b} that checks consistency of
a border coloring of a square. Then $s_{2N}=s_N+O(N)$, and so,
$s_N=O(N)$.

The proof of \ref {finitary}.\ref {fn-b} is essentially the same as the
proof of Theorem~\ref{bound} with two additions addressing minor issues.
 First, the self-stabilization counter of Section~\ref {stripepower}
restarts computation very rarely, with exponential intervals.
 Thus, a meaningless computation can run for a long time before the
re-initiation. It can fail to discover low complexity of the input, and
thus allow it in a large tiled square. This is easy to remedy just by
restricting the counters to (sufficiently large) polynomial values. Such
counters are implemented by a constant number of unary integers.

The second issue is that $x$ may appear near the border of large
$N\times N$ squares. Formally speaking, Proposition~\ref
{self-similarity-proposition} says nothing about finite tilings.
However, its proof (see~\cite {levin04,dls-mathint}) guarantees that
rank $k$ structure can be distorted only near the border of the tiled
region: for each $k$ there exists some $m(k)=\ooo(2^k)$ such that the
part of tiling that is $m(k)$ tiles apart from the border, has a correct
structure at rank $k$ (i.e., for $2^k\times 2^k$ squares). This can be
proved by induction over $k$ since the argument in~\cite {dls-mathint}
(that shows that $1$-tiles are grouped into $2$-tiles in a regular way)
uses only a small neighborhood.

So the checking goes on in the internal part of the tiled region
(square) and guarantees that the bit sequences there have no simple
substrings whose simplicity can be established fast. The problem is to
bring these complex bits to the border of the tiled region. This can be
easily done with the following trick: let us overlap four constructions
of the described type that propagate bits along the lines with direction
$(2,1)$, $(2,-1)$, $(1,2)$, $(1,-2)$ instead of vertical lines that we
have used for bit propagation. For every point on the border at least
one of these four directions brings us in the internal part of the tiled
region (square), so the bits on the border are also complex.

\end {proof}

\newpage\section {Turing Degrees of Tilings}\label {sec:recurs}

In this section we are concerned with Turing-degrees of tilings; we
provide a simple direct proof of the following theorem.

\begin {proposition}\label {recurs}
For each palette $\tau$ and for every undecidable set $A$ there exists a
$\tau$-tiling $T$ such that $A$ is not Turing-reducible to $T$.
 \end {proposition}

\begin {corollary}
For every palette $\tau$, there exists a $\tau$-tiling $T$ such that $T$
is not $\textbf{0}'$-hard. \end {corollary}

\begin {proof}
Let us consider the space $\mathcal{T}$ of all configurations made of
$\tau$-tiles (both tilings and configurations with tiling errors). This
space can be considered as $\tau^{\zed^{2}}$ and thus endowed with the
product (Cantor) topology. The set $\tau$ of local rules defines a
closed (compact) subset~$C$ of $\mathcal{T}$ consisting of all tilings.
This subset is an effectively closed subset (its complement is a union
of an enumerable family of basic open sets that correspond to violations
of local rules). By our assumption $C\ne\emptyset$.

Let $M$ be some oracle machine that uses a tiling $T\in\mathcal{T}$ as
an oracle. Let $x$ be some input for $M$ and $a$ be some output value
for $M$ (i.e., $a=0$ or $a=1$ if we consider machines that decide some
set). Consider the set of all oracles $T\in\mathcal{T}$ such that $M$
using $T$ produces answer $a$ on $x$. This set depends on $M$, $x$ and
$a$; we denote it by $U(M,x,a)$. It is easy to see that $U(M,x,a)$ is an
effectively open set: we have to simulate $M$ on input $x$ for all
possible oracles and look for all computation branches that end with
answer $a$. In this way we generate basic open sets whose union is
$U(M,x,a)$.

If $C \subset U(M,x,a)$, then $M$ produces answer $a$ on input $x$ for
all $\tau$-tilings. The crucial observation: if it is the case, we can
find it out eventually. Indeed, in this case the enumerable family of
base open sets that form $U(M,x,a)$ and the enumerable family of base
open sets that form $\mathcal{T}\setminus C$ together form a covering of
compact space $\mathcal{T}$. For compactness reasons, finite number of
sets are enough. Enumerating both families, we will discover this finite
covering at some point.

Now we can prove that there is no machine $M$ that reduces the
undecidable set $A$ to every $\tau$-tiling. (This statement is a weak
form of our theorem.) Indeed, this means that correct answer (we denote
it by $A(x)$) is produced for every input $x$ and every oracle $T\in C$,
\ie $C\subset U(M,x,A(x))$ for all $x$. But then we can compute $A(x)$
without oracle by looking for all $a$ such that $C\subset U(M,x,a)$. The
correct value $A(x)$ will be found; no other one can appear since
$U(M,x,a)$ and $U(M,x,a')$ are disjoint when $a\ne a'$.

The next step is to use diagonal argument and find $\tau$-tiling $T$
such that no machine decides $A$ using $T$ as an oracle. Let
$M_1,M_2,\dots$ be the enumeration of all (oracle) machines. We
construct a sequence $$C_0\supset C_1 \supset C_2\supset\ldots$$ of
effectively closed sets such that $C_0=C$, all $C_i$ are non-empty and
$M_i$ does not reduce $A$ to any of the oracles in $C_i$. Then the
intersection of all $C_i$ is non-empty because of compactness; every its
element $T$ is a $\tau$-tiling (since $C_0=C$) and no machine reduces
$A$ to $T$.

Assume that $C_{i-1}$ is already constructed. There are two
possibilities:

(1) If $C_{i-1} \subset U(M,x,A(x))$ for all $x$, then $A(x)$ is
computable for the same reason as before (where we had $C$ instead of
$C_{i-1}$); to compute $A(x)$ without oracle we look for $a$ such that
$C_{i-1} \subset U(M,x,a)$.

(2) If $C_{i-1}$ is not a subset of $U(M,x,A(x))$ for some $x$, then
choose some $x$ with this property and let $$C_i=C_{i-1} \setminus
U(M,x,A(x))$$ Then $C_i$ is non-empty, effectively closed and $M$ does
not produce correct answer $A(x)$ with input $x$ and any oracle $T\in
C_i$.

Note that the construction of $C_i$ is not effective (the choice of $x$
is not effective) but this is not needed: the only thing we need is that
each $C_i$ is effectively closed (though not uniformly in $i$). \end
{proof}

\newpage\bibliographystyle {abbrv}  \end {document}